# Topographically anisotropic photonics for broadband integrated polarization diversity

Jeff Chiles[1,*], Tracy Sjaardema[1], Ashutosh Rao[1] and Sasan Fathpour[1,2]

[1]CREOL, The College of Optics and Photonics, University of Central Florida, Orlando, Florida 32816, USA

[2]Department of Electrical Engineering and Computer Science, University of Central Florida, Orlando, Florida 32816, USA

[*]currently at the National Institute of Standards and Technology, Boulder, Colorado 80305, USA

**Integrated polarimetric receivers have the potential to define a new generation of lightweight, high-performance instrumentation for remote sensing. To date, on-chip polarization-selective devices such as polarizing beam-splitters have yet to even approach the necessary performance, due to fundamental design limitations. Here, we propose, simulate and experimentally demonstrate a method for realizing spatially-mapped birefringence onto integrated photonic circuits, deemed *topographically anisotropic photonics*. With this robust and widely tolerant approach, devices can be constructed with strongly polarization-dependent modal properties and minimal wavelength dependence. An integrated polarizing beam-splitter (PBS) is realized with unprecedented performance: a record 0.52 octaves of fractional bandwidth (116 THz), maximum insertion loss of 1.4 ± 0.8 dB, and a minimum extinction ratio of 16 ± 3 dB, pushing it into the realm of wideband spectroscopy and imaging applications. Additionally, novel photonic structures such as polarization-selective beam-taps and polarization-selective microring resonators are demonstrated, enabling new on-chip polarimetric receiver architectures.**

Polarization-diverse transmitter and receiver architectures are becoming increasingly prevalent in modern integrated photonic systems. A variety of applications benefit directly from this approach; the most well-established being high-bandwidth telecommunications[1]. Polarization diversity potentially enables a doubling of data bandwidth for a modest increase in chip size and little additional complexity. Indeed, this immediate impact has driven much of recent research in high-performance integrated components for polarization management on commercially popular platforms such as silicon-on-insulator (SOI). Crucially, the insertion loss of polarization management components must be kept low over a wide optical bandwidth. Typical components for this purpose include polarizers, polarizing beam-splitters (PBS) and polarization-splitter-rotators (PSR). Polarization diversity also has significant benefits for photonic systems targeted at sensing and imaging applications. Optical coherence tomography (OCT)[2] enables sub-surface imaging of translucent materials, such as organic tissue or fluids with exceptional resolution, and it has recently benefited from integrated photonic receiver technology allowing drastic reduction of the system bulk and complexity[3,4]. Polarization-diverse OCT receivers enable better negation of polarization artifacts and improved sensitivity in media with polarization-dependent scattering properties.

Another application for polarization-diverse integrated photonics is in remote spectroscopy. Remote spectroscopy is a metrology technique applicable to many fields of study, such as environmental monitoring, satellite reconnaissance and planetary science. Images that usually contain two dimensional intensity data, as well as some degree of spectral discrimination, are collected, which help to properly identify the composition of objects present in the scene. The inclusion of polarization data elucidates many details, and the combined study of spectral and polarimetric information in a scene is referred to as spectropolarimetric imaging[5]. Polarimetric light detection and ranging (LIDAR), leveraging dual lasers at 532 and 1064 nm wavelengths, has been used in the study of the vegetation canopy, the structure of which has significant implications in the study of microclimates, as well as large-scale climate change[6]. In atmospheric studies, optical polarimetry allows the precise determination of the phase state of clouds due to the polarization-dependent scattering of water crystals; the POLDER spaceborne instrument successfully conducted such measurements over a spectral band from 443 to 865 nm[7].

Currently, remote sensing systems rely on bulk or free-space optics. This leads to poor environmental stability and large size-weight and power requirements. Integrated photonics provides a means of drastically reducing these requirements and additionally driving down the cost to produce the overall system. For these reasons, the use of integrated photonics is already planned into upcoming technology for space exploration, such as NASA's ILLUMA laser modem[8].

However, to implement an integrated polarimetric remote sensing system, several key functions and performance requirements must be met. It must be capable of first separating the transverse electric (TE) and transverse magnetic (TM) signals. Additionally, it must achieve this over a broad bandwidth, typically on the order of an octave in fractional bandwidth[6,7]. Finally, it must do so without incurring excess insertion loss, which would degrade the sensitivity of the overall system. The performance of several state-of-the-art, experimentally demonstrated polarization-selective integrated devices is summarized in Table 1, emphasizing those with broad bandwidths. A more general review of polarization-management approaches is available in Ref. [9].

**Table 1: State-of-the-art broadband integrated polarization-selective devices**

| Work | Device | Bandwidth (nm) | Max. insertion loss (dB) | Min. ER (dB) |
| --- | --- | --- | --- | --- |
| Tan et al. [10] | PSR | 100 | ~2 | ~18 |
| Kim et al. [11] | PBS | 100 | 2.1 | 22.5 |
| Su et al. [12] | PBS | 150 | 3.4 | 10 |
| Xiong et al. [13] | TE-pass polarizer | 110 | 0.4 | 30 |
| Shahin et al. [14] | TM-pass polarizer | 110 | 2.4 | 18 |

All such devices take advantage of the dissimilar electric field distribution inherent to the TE and TM modes of the waveguides involved, which in turn leads to modal birefringence. Consequently, their cross-sections are specifically engineered to enhance this effect. An interesting approach, leveraging such modal birefringence in hybridized silicon nitride and phosphosilicate glass waveguide cores, was demonstrated in 1990[15]. A stronger differentiation between polarizations can be obtained by utilizing subwavelength structures (as in Refs. [13,14]) or tightly confined modes with significant field penetration outside the core. It is clear that, while acceptable insertion losses have been achieved using this general approach, the maximum bandwidth achieved is narrow, and hardly varies. We now consider the fundamental factors affecting this. The limitation in bandwidth essentially arises from the wavelength dependence of the field distribution, resulting in rapid changes in the weak-confinement regime, particularly in the case of TM modes in standard 220-nm-thick SOI. Furthermore, at shorter wavelengths, the TE and TM mode fields become more similar in their distribution as they reach greater confinement, which leads to reduced polarization selectivity. Finally, silicon itself exhibits absorption for $\lambda < 1.1$ μm, limiting its applicability in the short-wave and visible spectra.

It is apparent that these approaches, developed with telecommunications bandwidth requirements, are not suitable for wideband remote sensing applications. Most notably, achieving process compatibility with standard silicon photonics technology has been a major consideration. However, for the discussed high-performance applications, high-volume and low-cost production are not critical requirements, allowing leeway in the pursuit of more advanced fabrication techniques. Some other method of polarization management is required, particularly one that exploits an effect that is not strongly wavelength-dependent.

## Results

**Proposed scheme for on-chip polarization management.** The essence of the proposed scheme is manipulating the anisotropy of the waveguiding material itself, rather than manipulating its modal birefringence. Consider a waveguide with a core consisting of some uniaxial anisotropic material with the extraordinary axis pointing in the vertical direction (perpendicular to the plane of the substrate). The TE and TM modes of this waveguide experience a birefringence. Importantly, this birefringence remains even as the wavelength becomes shorter and the mode is strongly confined, in contrast to the behavior of isotropic-core waveguides, where they become less birefringent. However, making waveguides solely out of this anisotropic material does not directly enable broadband, polarization-selective devices, since the birefringence cannot be controlled or varied to any end. If instead the anisotropy could be adjusted spatially across the chip, or simply "turned off" in some waveguides and not in others, it could then be leveraged. This concept resembles selective disordering of quantum wells to tune the anisotropy in semiconductor waveguides[16,17]. By controlling the spatial distribution of material anisotropy on the chip, one can design devices that exploit transitions between different materials. This was later used to demonstrate a PBS[18]. The variable anisotropy concept has also been explored by a limited number of other groups using various approaches such as polymers[19,20] and organic crystals[21]. No broadband polarization-selective devices have been achieved with these approaches, which suffer from poor compactness, fabrication complexities or limited environmental stability due to the choice of materials.

These spatially-engineered anisotropic surfaces provide a means to achieve integrated polarization-selective devices with minimal wavelength sensitivity. But an approach is needed which puts this technology in line with the compactness, performance and robustness achieved by high-contrast photonic platforms. The requirements are: built from durable, stable materials; employs deposition methods compatible with conventional photonics fabrication techniques; includes the ability to co-integrate both anisotropic and isotropic films to achieve maximal contrast; and uses high-contrast waveguides with vertically symmetric cross sections. The last requirement is subtle but still crucial, as it avoids the transition to cut-off that would be inevitable for any asymmetric waveguide geometry; this is a limitation for all of the prior methods for engineered anisotropy, as well as for approaches that effectively augment conventional waveguides with anisotropic overlayers[22,23].

We propose a novel approach called *topographically anisotropic photonics* (TAP) to achieving all these requirements by employing a combination of the photonic damascene process[24] with precisely engineered dielectric film stacks. The material anisotropy is derived from a multilayer stack (MLS) consisting of alternating layers (deep-sub-wavelength thickness, $t_L$ and $t_H$) of, for example, silicon dioxide ($SiO_2$) (*L*, low index) and silicon nitride ($Si_3N_4$) (*H*, high index), exhibiting form-birefringence in the refractive index as described below by the effective medium theory[25]:

$$n_{eff,TE}^2 = fn_H^2 + (1-f)n_L^2 \tag{1}$$

$$\frac{1}{n_{eff,TM}^2} = \frac{f}{n_H^2} + \frac{(1-f)}{n_L^2} \tag{2}$$

$$t_{H,L} \ll \lambda_H \tag{3}$$

where *n* is the refractive index, and *f* is the fill fraction of the high-index material in the MLS. This is a valid approximation when the thickness of the layers is much smaller than the wavelength of light inside the medium of high index (equation 3). Birefringence is observed in the material refractive index due to the different behavior of TE- and TM-polarized waves. Such so-called "metamaterial" stacks have been exploited for their anisotropic properties to modify the behavior of evanescent fields of waveguide claddings[26].

Now imagine a waveguide (Fig. 1a) with a silicon oxynitride (SiON) core and an MLS cladding with refractive indices, as indicated in the figure, satisfying the following conditions:

$$n_{core} > n_{eff,TM}, \tag{4}$$

$$n_{core} < n_{eff,TE}, \text{ and} \tag{5}$$

$$n_{clad} < n_{core}. \tag{6}$$

If the MLS cladding material is constructed such that equations 4-6 are satisfied through equations 1 and 2, the waveguide displays intrinsic single-polarization behavior, in that TE-polarized light will become anti-guided due to the negative index contrast, and TM-polarized light will be guided normally. Similarly, if the core and cladding materials are simply swapped in place such that the core is MLS and the cladding is the isotropic material with $n_{core}$ (Fig. 1b), the waveguide will only guide TE-polarized light.

The proposed approach indeed allows both of these arrangements, and more, to be achieved in the same process, as follows. The MLS is first deposited on an oxidized silicon substrate. Next, the anisotropy is "mapped" on the surface by etching the MLS in some areas and then filling it with SiON or $SiO_2$ as core or cladding materials. Etch-back techniques are used to restore surface flatness after the etching and refill deposition steps. A complete description of the fabrication technique is provided in Methods. A variety of unique waveguiding "states" may be realized with TAP, which are summarized in Fig. 1.

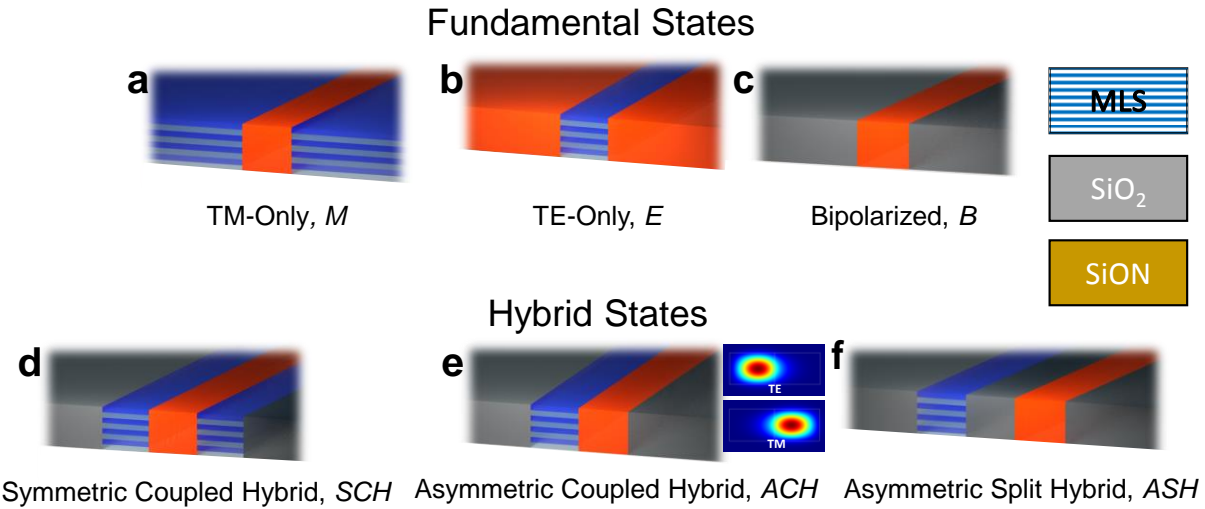

**Figure 1 | Fundamental and hybrid waveguide arrangements used in topographically anisotropic photonics.** In (e), the inset to the right shows simulated intensity distributions for TE- and TM-polarized light for the *ACH* state, illustrating how each polarization-mode is tightly confined to the respective region of greater effective index.

The states are divided into fundamental and hybrid types. The fundamental states include TE-only (*E*), TM-only (*M*) and bipolarized (*B*). They consist of one core material with symmetric claddings of another material. The hybrid states, which each use two different core materials, include the symmetric coupled hybrid (*SCH*), asymmetric coupled hybrid (*ACH*) and asymmetric split hybrid (*ASH*). In each fundamental and hybrid state, a different relation exists between the TE and TM modes. For example, in the *ACH* hybrid state, the TE mode is concentrated in the MLS core due to the higher refractive index, whereas the TM mode is concentrated in the SiON core, due to the low refractive index it experiences in the MLS (Fig. 1e). Importantly, the polarization-selective behavior of the hybrid states is robust to many factors, including confinement factor, waveguide geometry tolerances, and even some degree of variability in the refractive indices.

**Design approach.** Now, we consider how to apply TAP to the goal of achieving broadband and efficient polarization-selective devices. With TAP, light can start in any one of the fundamental states, generally in *B* for incoming light of an indeterminate polarization. Next, to parse the TE and TM modes, a transition may be made to one or more of the hybrid states, which will result in a new polarization-mode distribution. Finally, it undergoes a transition back to another fundamental state to resume transmission elsewhere on the chip. In the following sections, devices such as PBS and polarization-selective beam-taps and microring resonators are designed and simulated using this simple approach.

**Polarizing beam splitter design and simulations.** The approach for designing a PBS is now detailed (Fig. 2). It is assumed that the input state is *B*, such that it is bipolarized. Next, a transition is made to the hybrid *ACH* state mentioned earlier (Fig. 1e), which forces the TE and TM modes into separate core regions. Afterward, a transition is made to the *ASH* state by introducing a "wedge" of silicon dioxide cladding in between the two cores. The gap between the SiON and MLS cores is gradually widened until they do not interact. Finally, at the output, the MLS core in the TE arm is replaced with SiON (not pictured) to be compatible with ordinary bus waveguides on the chip. Thus, the output state is again *B* for both the TE and TM arms, with the two polarization modes efficiently diverted into their respective paths.

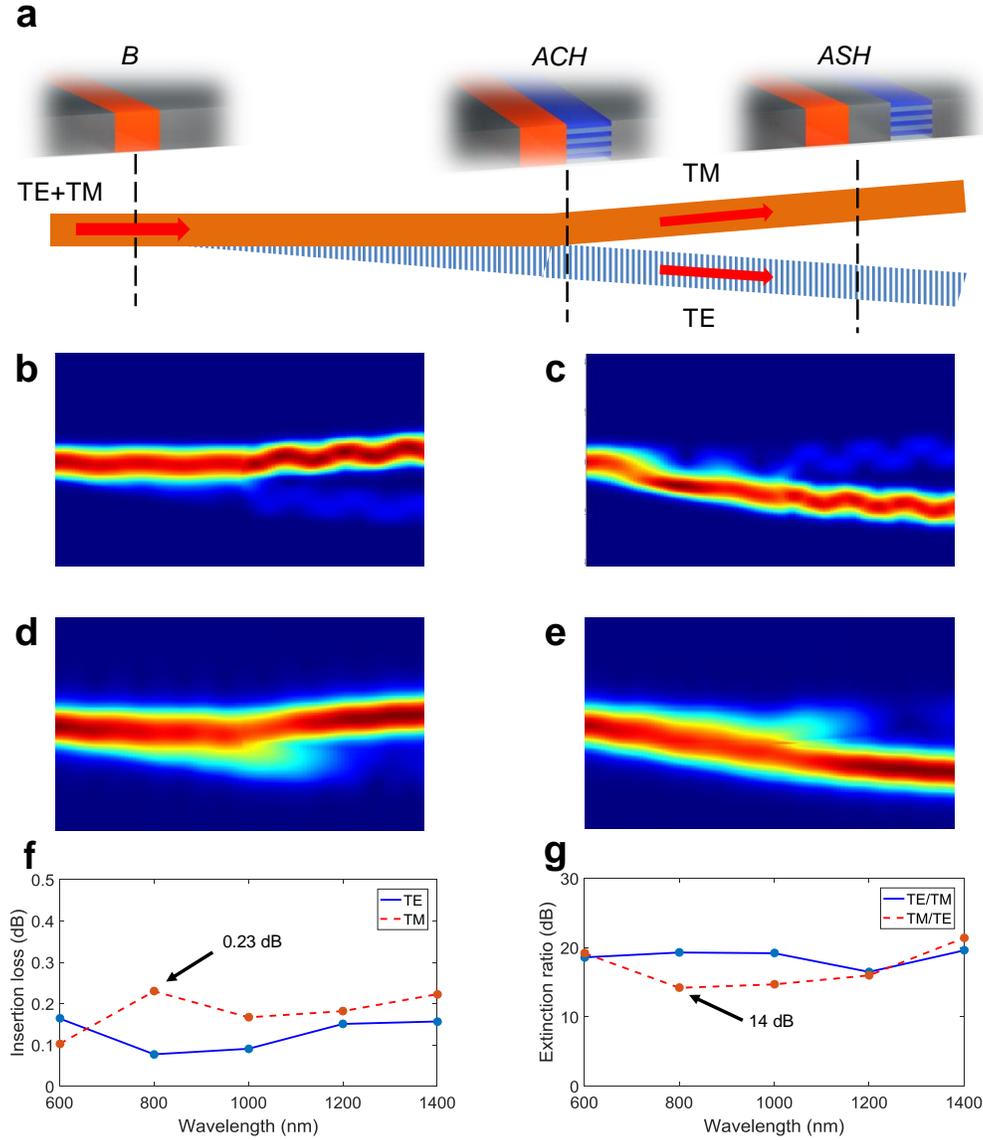

**Figure 2 | Polarizing beam splitter design and simulations. (a)** Design employing state transitions (whitespace is SiO$_2$ cladding). The *B--ACH* transition length is 60 μm, the full wedge angle for the *ACH--ASH* transition is 1°, and the maximum width of the SiON and MLS cores is 1.4 μm, each. The PBS is ~121 x 5 μm$^2$, including the output wedge, not including waveguide width-adjustment tapers. **(b-d)** Simulated normalized electric field (top-view) of the PBS for **(b,c)** λ = 600 nm, **(d,e)** λ = 1400 nm; **(b,d)** TM-polarized input, and **(c,e)** TE-polarized input. **(f)** Simulated insertion loss spectrum; **(g)** simulated ER spectrum.

The proposed PBS design is simulated in Lumerical MODE Solutions™ using the eigenmode expansion (EME) algorithm, incorporating an effective anisotropic material for the MLS to reduce meshing requirements. The material dispersion is relatively small for these materials and is thus neglected for the structural simulations in this work. Simulations are conducted by injecting TE or TM light into the *B*-type input port and monitoring the extinction ratio (ER) and insertion loss at the TE and TM output ports. The performance is examined at λ = 600 – 1400 nm. The normalized electric field plots are shown in Fig. 2b-e for all cases, and the performance is summarized in Fig. 2f,g. Across the span from λ = 600 to 1400 nm, it is observed that the insertion loss remains below 0.3 dB and the ER is always greater than 14 dB. Some multimode oscillation is visible on the output ports at shorter wavelengths, though this may be readily addressed in the future by implementing more refined splitter geometries incorporating curved *Y*-junctions and parabolic widening profiles. However, it is quite remarkable that the design continues to work with high splitting efficiency even in the highly multimode regime at λ = 600 nm. This would not be possible for conventional PBS designs that rely on precise mode shape engineering with purely isotropic materials. This behavior could be highly useful for remote-sensing receiver chips where higher power collection efficiency is desirable and multimode operation is tolerable.

**Polarization-selective beam-taps and microring resonator design and simulations.** Consider a microring resonator, in which light is coupled into the ring cavity, with the amount coupled dependent on the coupling coefficient. For applications such as polarimetric remote spectroscopy, it may be desired to have light of only one polarization resonate in the ring, so as to apply a spectral filtering profile to that particular polarization state. The other polarization ideally would not couple into the ring at all. We refer to this device as a "polarization-selective microring resonator" (PSMR). A simple way to achieve a similar effect might be to simply pass the incoming light through a PBS, and insert the microring resonator into one arm with the desired polarization, then recombine the polarization channels through another PBS. Despite its simplicity, this approach is inefficient, consumes significant chip area and adds unnecessary insertion losses from the additional components.

TAP provides a means of realizing broadband, low-loss PSMRs for the first time. The design of a TAP PSMR (shown schematically in Fig. 3a) follows a similar procedure to that of the PBS. For a microring resonator, the coupling region is nothing more than a beam-tap engineered to give a specific (and usually small) coupling coefficient. Thus, one can imagine a TAP PBS that is instead engineered to be a polarization-selective beam-tap with a small but finite coupling coefficient for the resonated polarization, and zero coupling for the "through-polarization." As with a normal beam-tap, the degree of coupling is tunable by the gap between the bus waveguide and the drop waveguide. This can be achieved in a *B -- ASH -- B* transition, as the bus waveguide crosses the ring cavity.

As with the PBS, the polarization splitting is achieved by using different core materials for the TE and TM ports. For a TE-selective PSMR, the core material in the ring waveguide should thus be MLS, and the core material in the bus should be SiON. Also, the transition to and from the *ASH* state should allow the mode to gradually evolve. This follows from the use of a gradually widened MLS core in the previous PBS design. Since there is now a gap of $SiO_2$ cladding separating the bus and the ring waveguides in the PSMR, this can instead be implemented by gradually tapering down and up the width of the ring waveguide. The degree of tapering and the gap allow control over the polarization-dependent coupling ratio (PDCR) and the power coupling coefficient, respectively. The PDCR is defined simply as the ratio of power coupling coefficient into the tap-off port (or into the ring cavity, alternatively) for one polarization state, divided by the power coupling coefficient into the tap-off port for the other polarization state. In a "TE-selective" PSMR, for instance, the ratio would be $P_{TE}/P_{TM}$, where $P$ is the power coupling coefficient into the tap-off port.

Simulations are conducted at λ = 850 nm for TE- and TM-polarized incoming light, for both the TE-selective and TM-selective PSMR cases. Additionally, the case of a purely isotropic core material (SiON) in both the ring and bus waveguides is examined to provide a reference for the default behavior. The simulation is limited to the beam-tap (coupling section), rather than the complete ring structure, to reduce memory requirements. The two-dimensional normalized electric field profiles of all cases are shown in Fig. 3b-d, and the simulated polarization-selective beam-tap (PSBT) tap-port coupling coefficients and PDCRs are summarized in Table 2.

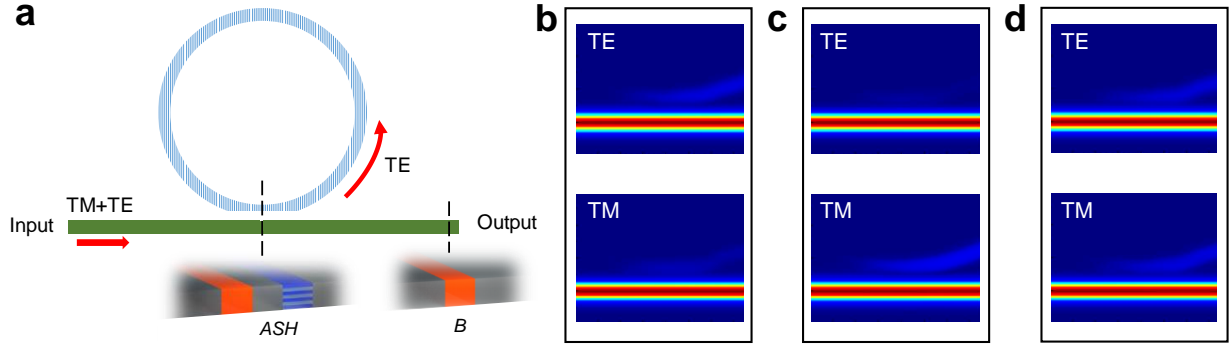

**Figure 3 | Polarization-selective beam-tap/microring-resonator design and simulations. (a)** Schematic top view of the proposed design approach for a TE-selective microring resonator. The bus and ring waveguide nominal widths are 0.9 μm, the ring radius is 100 μm, and a width taper-down to 700 nm is applied in the ring waveguide by "clipping" the bottom portion. An edge-to-edge gap of 700 nm was used in the clipped waveguide region. Whitespace is considered to be $SiO_2$ cladding. **(b-d)** Simulated normalized electric field profile from a top-view of beam-taps for the **(b)** TE-selective design, **(c)** TM-selective design, and **(d)** SiON-core-only design (non-selective).

**Table 2: Simulated polarization-selective beam-tap performance**

| Device type | TE coupling coefficient | TM coupling coefficient | PDCR, dB |
|---|---|---|---|
| TE-selective | 0.32% | 0.035% | 9.7 |
| TM-selective | 0.005% | 0.72% | 21 |
| Nonselective | 0.28% | 0.35% | 1.0 |

For both the TE- and TM-selective designs, wherein the only difference is simply swapping the position of the MLS and SiON core materials, a high PDCR of up to 21 dB is attainable. Comparatively, when the same ring is simulated but with SiON in both the ring and bus cores (the reference example), there is only a small PDCR of 1.0 dB. This validates the design approach of using an anisotropic index contrast between the core materials of the bus and ring waveguides to maximize the PDCR in this device. In some ways, this device's engineered control over coupling coefficients resembles the recent use of deep-subwavelength structures with a different goal of increasing the density of nanophotonic components[27].

**TAP sample preparation.** TAP samples are prepared for testing of the PBS and PSMR designs, as follows. An MLS consisting of 15 pairs of $Si_3N_4/SiO_2$ layers (Fig. 4) is optimized in fill fraction to give a TE refractive index of ~ 1.71 and a TM refractive index of 1.65. The total thickness of the MLS is approximately 700 nm. Prism-coupling measurements show intrinsic MLS material loss of 4-5 dB/cm at $\lambda$ = 633 nm and 0.8 dB/cm at $\lambda$ = 827 nm for TE-polarized light. Next, the SiON film deposition is optimized to give a refractive index of ~ 1.68. It also showed low material loss (< 1 dB/cm at $\lambda$ = 633 nm). The MLS is deposited on an oxidized silicon substrate. Devices are fabricated by patterning and etching trenches to define regions where SiON or $SiO_2$ are filled in. The full fabrication process is described in Methods.

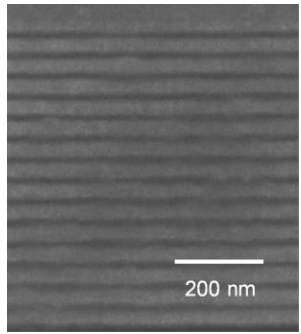

**Figure 4 | Scanning-electron micrograph (SEM) of the fabricated MLS cross-section.** The apparent roughness is a result of wet-etching performed to enhance the visible contrast of the layers.

**Polarizing beam splitter characterization.** A chip layout is created to allow robust testing of the earlier proposed PBS design. An example propagation path for a PBS test device is given in Fig. 5a. A common input waveguide 1.5 μm-wide is first tapered down to 0.7 μm and propagated for 300 μm to extinguish most higher-order mode content. Then, it passes through a 50:50 *Y*-junction splitter into a "device" and a "reference" path. In the device path, the waveguide width is tapered to the appropriate value for the PBS, and then back down to 0.7 μm afterward to allow a larger separation between the TE and TM ports. Finally, it is tapered up to 1.5 μm for low-loss propagation to the output ports. In the reference arm, the waveguide is tapered up to the 1.5 μm output width after a short propagation length to provide appropriate path length matching between the reference and the device path.

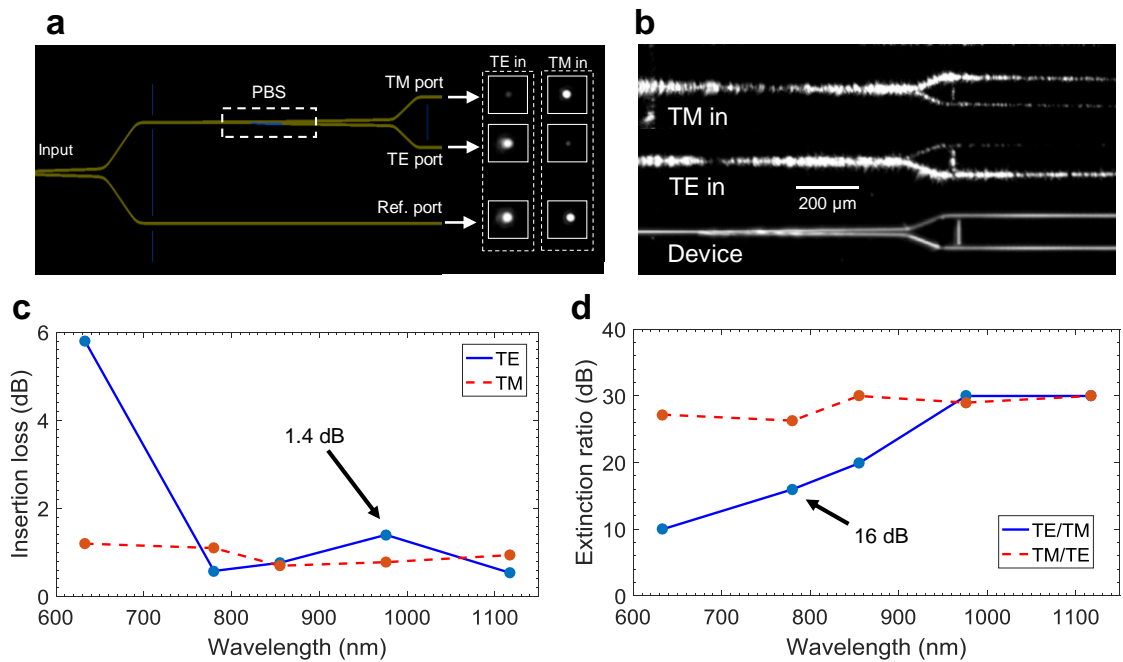

**Figure 5 | Device layout and characterization results for polarizing beam-splitter.** **(a)** Top-view schematic of the device layout for testing the PBS, showing the common input bus splitting into the reference and PBS output paths. An example measurement of a port power measurement via top-view imaging is also included (λ = 855 nm). **(b)** Top-view from a camera showing scattered light from the PBS arms as TE and TM input polarizations are selected; **(c)** Experimental insertion loss spectrum observed from the PBS with maximum in-band loss highlighted; **(d)** Experimental ER spectrum, with minimum in-band value highlighted.

This scheme allows highly repeatable measurements, largely independent of the input coupling conditions. The PBS design is fabricated with the same dimensions simulated earlier (see Methods). The chip is partially diced and then cleaved on the input side to provide a smooth facet for coupling. On the output side, the waveguides are not cleaved, but terminated in etched facets at the output to provide highly consistent out-coupling for all ports. For characterization, light is coupled in through a 20X objective lens from free-space laser sources at $\lambda$ = 633, 780, 855, 976 and 1,117 nm. The input light is polarized immediately after the source with a Glan-Taylor polarizer. An optical micrograph of a fabricated PBS and the fan-out from the device is shown in Fig. 5b. Light scattering from the output facets is imaged from a top view with a digital camera, such that the TE, TM and reference ports are all visible. The optical power in each port is thus proportional to the integral of the intensity profile of the scattered light from each port (Fig. 5a). For calculation of insertion losses, the brightness is always adjusted to avoid saturation of the pixels. Several measurements are conducted when necessary and averaged to obtain the final results. The uncertainty in all PBS measurements is ± 0.8 dB for insertion losses and ± 3 dB for ERs. The characterization results are summarized in Fig. 5c,d. A maximum distinguishable ER value of 30 dB is inferred from measurements. Insertion losses through the PBS device are reported as excess losses compared to those experienced in the reference waveguide path.

The fabricated PBS achieved ER > 16 dB and maximum insertion losses of 1.4 ± 0.8 dB over a 337 nm span of $\lambda$ = 780 – 1,117 nm, representing a fractional bandwidth of 0.52 octaves. A minimum ER (considering both TE/TM and TM/TE) of 16 ± 3 dB is achieved over the same bandwidth. Furthermore, the ER exceeds 25 dB at longer wavelengths. This is the broadest operational bandwidth in an integrated PBS ever demonstrated, by more than a factor of 3 (in octave space), while achieving maximum insertion losses more than a factor of 2 lower, and a 6 dB minimum ER improvement[12].

While the experimental results show drastic improvements over all prior integrated PBS devices, there is still a discrepancy between these and the simulated performance, which predicted lower losses of approximately 0.3 dB. One limitation is proximity effects during etching of nanoscale features such as the wedge splitter. For very narrow gaps, the MLS or SiON may not be etched fully, resulting in a deviation from the ideal geometry. Specifically, this reduces the effective contrast achieved between the MLS, $SiO_2$ and SiON zones. This could be addressed in the future by optimizing the etching recipe. Another issue could be the lack of perfectly flattened surfaces after the etch-back-based planarization. A local surface planarization (in the vicinity of etched features) of approximately 200 nm is achieved, which is large enough to influence the index contrasts and interfere with ideal performance. This could be mitigated with chemical-mechanical polishing (CMP) in the future.

**Polarization-selective beam-tap and microring resonator characterization.** Another TAP chip is fabricated to assess the performance of the PSMR and PSBT designs proposed earlier. The PSBTs are tested by coupling in light at $\lambda$ = 633, 780 and 855 nm, using essentially the same approach as for the PBS, although an aspheric lens is employed for improved focusing on the waveguide facet. Since the coupling coefficient is designed to be small, multiple exposure lengths are required on the camera at the output to acquire sufficient dynamic range to accurately assess the performance. The gain curve of the sensor is calibrated to allow linear power measurements with different exposures. An example chip layout schematic is shown in Fig. 6a, where the common input is split into device and reference paths. In the device path, it is further split and diverted into a TE-selective and a TM-selective PSBT, each having a tap and a bus port. The tap power coupling coefficient is used for the PDCR measurement.

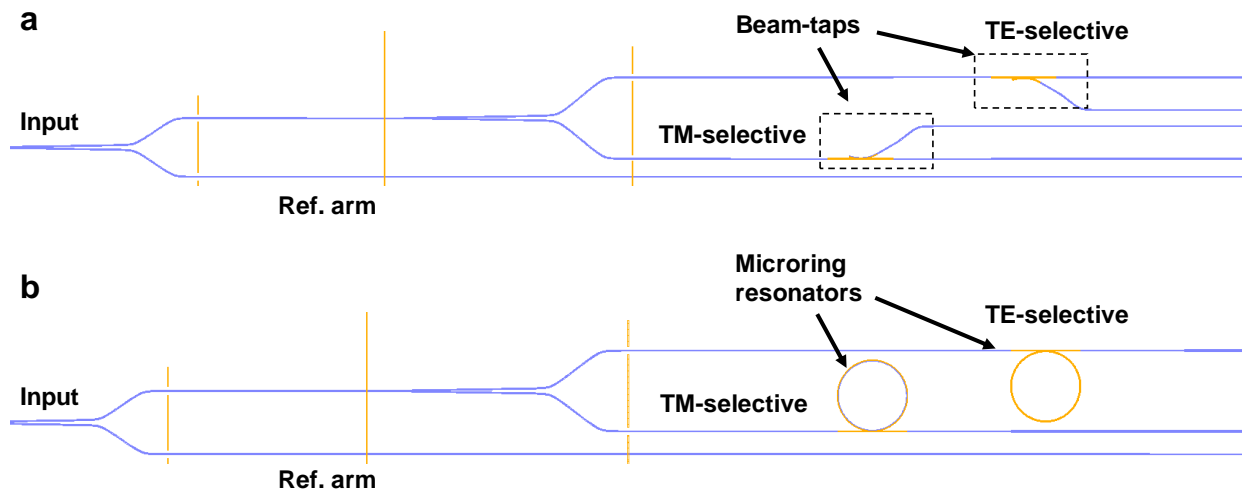

**Fig. 6 | Device layout for polarization-selective beam-taps and microring resonators.** (a) Beam-taps; (b) Microring resonators. The same geometrical parameters are applied to the TM- and TE-selective PSBTs in a given measurement path (i.e., waveguide width, coupling gap); only the composition of the bus and tap waveguide cores (SiON or MLS) are swapped as dictated by the design. Vertical lines are etched trenches to limit slab-coupled light.

**Table 3: Experimental polarization-selective beam-tap performance**

| Wavelength (nm) | Device type | TE coupling coefficient | TM coupling coefficient | PDCR, dB |
|---|---|---|---|---|
| 780 | TE-selective | 2.65% | 0.60% | 6.5 ± 1 |
|  | TM-selective | 0.09% | 2.22% | 14 ± 1 |
| 855 | TE-selective | 2.77% | 0.78% | 5.5 ± 1 |
|  | TM-selective | 0.21% | 1.83% | 9.4 ± 1 |

The data are summarized in Table 3. The best devices employed an edge-to-edge coupling gap of 600 nm at the middle of the coupler section. At $\lambda$ = 633 nm, no useful PDCR is obtained; this may be due to increased sensitivity to aspect-ratio dependent etching in the narrow gap. However, a PDCR > 5.5 ± 1 dB, and up to 14 ± 1 dB, is maintained at $\lambda$ = 780 and 855 nm. This measurement provides experimental confirmation that strong polarization dependence can be designed into beam-taps based on the TAP platform, through the exploitation of engineered anisotropy and modal state transitions.

Next, the TM- and TE-selective rings of 100-µm-radii are tested with a polarized single-longitudinal-mode vertical-cavity surface-emitting laser (VCSEL) source at $\lambda$ = 855 nm (Thorlabs CPS850V), oriented at 45° with respect to the plane of the chip. The ring resonators are identical in structure to the fabricated set of PSBTs, only that the tap port is closed on itself to form the resonant cavity (Fig. 6b). The input light is polarized in the horizontal and vertical directions as required with a Glan-Taylor polarizer. To sweep the laser through the ring resonances, it is thermally tuned from room temperature to 50°C; the thermistor voltage-to-temperature-to-wavelength characteristic curve is calibrated prior to measurements using an optical spectrum analyzer (OSA).

Light is coupled onto the chip via a 20X objective lens, and measured at the output in the same method used for the PBS, but using a time-lapse of images to accumulate the transmission spectrum. The transmission spectra through TM- and TE-selective PSMRs across a full sweep of the laser wavelength are plotted in Fig. 7a,b. When necessary, the spectra are corrected for slow drifts in the input power via a spline function to remove an overall tilt. Each transmission spectrum trace is fitted to the characteristic response of a notch resonance[28]. A peak loaded quality factor of 47,000 ± 5,000 is observed in the case of TM-polarized light into the TM-selective resonator. The uncertainty in this value originates mainly from a thermal-delay-induced deviation from the wavelength calibration.

For a designed coupling gap of 700 nm at the middle of the coupler section, the TM-selective PSMR with a waveguide width of 900 nm exhibits notch resonances with an ER of 11 dB for TM-polarized input light, and no visible resonance for TE-polarized light (Fig. 7a). These results demonstrate a strong polarization selectivity as expected from the design. The TE-selective PSMR shows resonances for both polarizations, but exhibits a strong ER contrast, $\Delta ER$, of 8.5 dB between TE- and TM-polarized input light (Fig. 7b). Indeed, these results are highly consistent with the reduced PDCR observed in the TE-selective PSBTs measured earlier in this section. By using TAP, strongly polarization-dependent resonant spectral manipulation is successfully implemented on an integrated platform, without requiring polarization-attenuating elements in the resonant cavity or any external PBS devices.

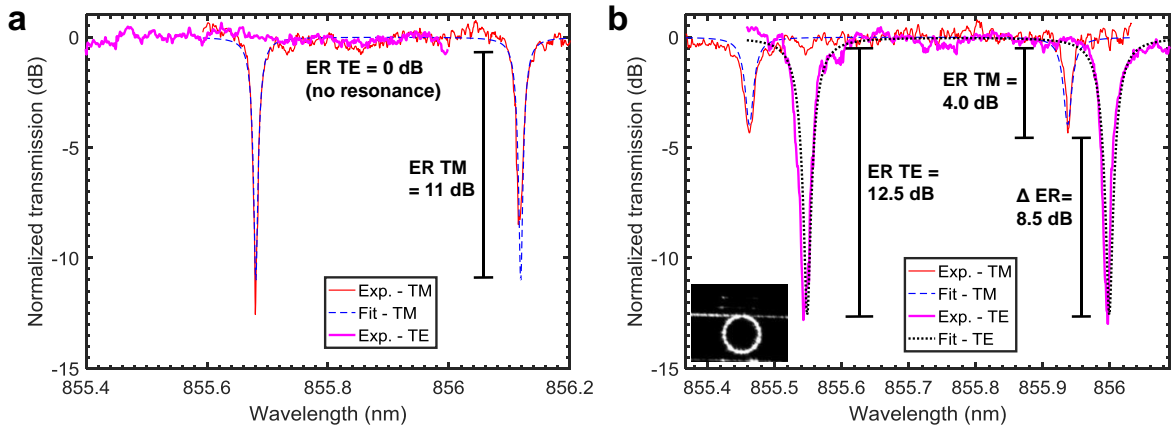

**Figure 7 | Experimental characterization results for polarization-selective microring resonators.** (a,b) Transmission spectrum through PSMRs. (a) TM-selective PSMR. (b) TE-selective PSMR, showing the resonances for TM- and TE-polarized input light and the relative ERs. Inset: top-view infrared micrograph of a PSMR on-resonance.

## Discussion

An approach to achieving broadband and highly efficient polarization-selective integrated photonics, called *topographically anisotropic photonics* (TAP), is proposed, simulated and validated with fabricated and characterized devices. At its core, TAP leverages a set of possible waveguiding configurations that are differentiated by spatially mapped optical anisotropy. By selecting an appropriate combination of states and employing smooth transitions between them, we show in this work that polarized light can be efficiently parsed over bandwidths that were previously inaccessible to integrated photonics. Because the birefringence is built into the core multilayer stack, TAP bypasses the need for high-precision geometrical tolerances, instead taking advantage of the naturally high precision achieved with modern chemical vapor deposition technology.

Polarizing beam splitters (PBS) are investigated using this approach. A record fractional bandwidth of 0.52 octaves or 116 THz is demonstrated, while maintaining a minimum ER of 16 ± 3 dB and a maximum insertion loss of 1.4 ± 0.8 dB, exceeding the prior record in integrated PBS bandwidth by more than a factor of three, while achieving significantly lower loss. This broad improvement in overall performance has immediate implications in polarimetric receiver systems, where integrated photonics can lead to cost and bulk reductions.

We also show how TAP can be used to enable novel photonic structures such as polarization-selective beam-taps and microring resonators. Such devices are designed using state-transitions in a similar way for the PBS. Beam-taps with coupling coefficients in the 1% range are shown to exhibit strong polarization selectivity up to 14 ± 1 dB in their coupling ratio. A TM-selective microring resonator is fabricated and characterized, showing deep notch resonances for TM-polarized light and completely suppressed resonance for TE-polarized light.

Further improvements to the planarization and etching are expected to drastically improve the performance in both loss and bandwidth. Future investigation will also include multi-element polarimetric receivers. This technology shows significant promise for ultra-broad bandwidth integrated polarization-diverse photonics for sensing and high-bandwidth communications.

## Methods

**TAP device fabrication.** The 700 nm-thick MLS core deposited on oxidized silicon acts as the initial device layer for TAP chips. First, electron-beam lithography is used to pattern a resist mask (ZEP 520, ZEON Corp.™) to define where SiON material will replace it. The MLS is fully etched with $CHF_3/SF_6$ chemistry. After resist stripping, the trenches are re-filled with a thick layer of ~ 1.8 µm SiON using plasma-enhanced chemical vapor deposition (PECVD). The uneven SiON film is planarized with 1 µm of photoresist (Microposit™ S1805), and then etched halfway through the SiON thickness with $CHF_3/O_2$ chemistry giving nearly 1:1 selectivity between the resist and SiON. The coating and etch-back are repeated, and the etch is terminated at the top of the MLS. Another lithography and pattern etching is conducted (using $CHF_3$ chemistry) to define trenches where $SiO_2$ cladding will be filled in. Afterward, the new trenches are refilled and the surface is top-cladded with PECVD $SiO_2$ to symmetrize the device layer.

## Acknowledgements

This work was supported by the National Science Foundation (NSF) CAREER (ECCS-1150672). We thank A. Dogariu and C. Constant for loaning the 1,117 nm laser source, neutral-density filters and polarizers. We also acknowledge S. Khan for assistance with the measurements.

## Author contributions

J.C. proposed the concept, performed the design and simulation, piloted the fabrication technique and experimental setup, and performed data analysis. T.S. fabricated the final samples, refined and expanded the experimental setup and data analysis, and performed the measurements. A.R. characterized the fabricated MLS film properties. S.F. proposed device concepts, analyzed data, and supervised the project. J.C. and S.F wrote the manuscript.